\documentclass[twocolumn,showpacs,preprintnumbers,amsmath,amssymb]{revtex4}
\usepackage{tipa}
\usepackage{amssymb}
\usepackage{txfonts}
\usepackage{hyperref}
\usepackage{graphicx}
\usepackage{epsfig}
\begin{document}

\title{Statefinder Diagnostic for Phantom Model with $V(\phi)=V_{0}\exp(-\lambda{\phi}^2)$}
\author{Baorong Chang \footnote{changbaorong@student.dlut.edu.cn, telephone:0411-84707869},
Hongya Liu\footnote{Corresponding author: hyliu@dlut.edu.cn,
telephone:0411-84707765},Lixin Xu  and Chengwu Zhang}
\affiliation{School of Physics \& Optoelectronic Technology, Dalian
University of Technology, Dalian, 116024, P. R. China}


\begin{abstract}

We investigate the phantom field with potential
$V(\phi)=V_{0}\exp(-\lambda{\phi}^2)$ and dark matter in the
spatially flat FRW model. It has been shown by numerical calculation
that there is a attractor solution in this model. We also apply the
statefinder diagnostic to this phantom model. It is shown that the
evolving trajectories of this scenario in the $s-r$ diagram is quite
different form other dark energy models.

\end{abstract}

\maketitle

Recently, the observations of high redshift type Ia
supernovae\cite{Ia} reveal the speeding up expansion of our universe
and the surveys of clusters of galaxies show that the density of
matter is very much less than the critical density\cite{SDSS}, and
the observations of Cosmic Microwave Background (CMB) anisotropies
indicate that the universe is flat and the total energy density is
very close to the critical one with $\Omega _{total}\simeq 1$
\cite{CMB}. These three tests nicely complement each other and
indicate that the dominated component of the present universe is
dark energy(DE). Dark energy occupies about $73\%$ of the energy of
our universe, while dark matter(DM) about $23\%$, and the usual
baryonic matter occupy about $4\%$. The accelerating expansion of
the present universe is attributed to the dark energy which is an
exotic component with negative pressure, such as the cosmological
constant $\Lambda$ \cite{sahni9904398,cosmological constant} with
equation of state $w=-1$, a dynamically evolving scalar field
(quintessence) \cite{quintessence1,quintessence2} with $w>-1$ or the
phantom \cite{phantom} with $w<-1$, meanwhile the accelerating
expansion of universe can also be obtained through modified
firedmann equation\cite{astro-ph/0403228} and brane
world\cite{brane}. The fine tuning problem is considered as one of
the most important issues for dark energy models and a good model
should limit the fine tuning as much as possible. The dynamical
attractors of the cosmological system have been employed to make the
late time behaviors of the model insensitive to the initial
conditions of the field and thus alleviate the fine tuning problem,
which has been studied in many quintessence models
\cite{quattractor1}. In phantom system, this problem has been
studied with cosine potential\cite{guozk}, exponential and inverse
power law potential in \cite{phattractor1,phattractor2}. In this
letter, we study the phantom field $\phi$ with the potential
$V(\phi)=V_{0}\exp(-\lambda{\phi}^2)$ and the dark matter in the
spatially flat FRW model. The late time behavior of the cosmological
equations will give accelerated expansion and a constant ratio
between dark matter energy density $\rho_{m}$ and phantom energy
density $\rho_{\phi}$. This behavior relies on the existence of an
attractor solution, and the late time cosmology insensitive to the
initial conditions for dark matter and dark energy. Therefore, the
fine tuning problem can be alleviate in this model.

In this paper, we will first show the attractor behavior in this
scenario and, then we perform a statefinder parameter diagnostic for
this model. The statefinder parameter introduced by Sahni et al.
\cite{statefinder} are proven to be useful tools to characterize and
differentiate between various dark energy models. We show in this
paper the evolving trajectory of the $s-r$ diagram is quite
different from this of other dark energy models.

We consider a universe model which contains phantom field $\phi$ and
the dark matter $\rho_{m}$. The Friedmann equation in a spatially
flat FRW metric can be written as
\begin{equation}
H^{2}=\frac{1}{3}(\rho_\phi+\rho_{m}),\label{H^2}
\end{equation}
the Planck normalization $M_{p}=1$ has been used here, $\rho_{m}$ is
the energy density of the dark matter, and the dark matter possesses
the equation of state $P_{m}=(\gamma_{m}-1)\rho_{m}$. The energy
density and pressure of the phantom field $\phi$ are $\rho_{\phi}$
and $P_{\phi}$, respectively,
\begin{equation}
\rho_\phi=-\frac{1}{2}{\dot{\phi}}^{2}+V(\phi),\label{density}
\end{equation}
\begin{equation}
P_\phi=-\frac{1}{2}{\dot{\phi}}^{2}-V(\phi).\label{pressure}
\end{equation}
where $V(\phi)$ is the phantom field potential,
$V(\phi)=V_{0}\exp(-\lambda{\phi}^2)$.

Since the energy of dark energy and dark matter is conserved
respectively, the equation of motion for DE and DM can be obtained:
\begin{equation}
\dot{\rho}_\phi+3H\rho_\phi(1+w_{\phi})=0,\label{DE}
\end{equation}
\begin{equation}
\dot{\rho}_{m}+3H\rho_{m}(1+w_{m})=0.\label{DM}
\end{equation}
where the parameter of equation of state for the phantom field is
given by:
\begin{equation}
w_{\phi}=\frac{-\frac{1}{2}{\dot{\phi}}^{2}-V(\phi)}{-\frac{1}{2}{\dot{\phi}}^{2}+V(\phi)}.
\end{equation}
and we can get the equation of motion for the scalar field $\phi$
\begin{equation}
\ddot{\phi}+3H\dot{\phi}+2\lambda{\phi}V(\phi)=0.\label{scalar}
\end{equation}

Using the Friedmann equation eq.(\ref{H^2}), the eq.(\ref{scalar})
can be written as
\begin{equation}
\phi^{''}H^{2}+[(1-\frac{1}{2}\gamma_{m})\rho_{m}+V(\phi)]\phi^{'}=-2\lambda{\phi}V(\phi).\label{u}
\end{equation}
where
\begin{equation}
H^{2}=\frac{1}{3}\frac{V(\phi)+\rho_{m}}{1+\frac{1}{6}{\phi^{'}}^{2}},\label{H^22}
\end{equation}
primes denote derivatives with respect to $u=\ln(a/a_0)=-\ln(1+z)$,
where $z$ is the redshift, and $a_0$ represents the current scalar
factor.

It is known that the attractor solution can be found analytically in
the exponential potential. However, in our case, it is difficult to
obtain the analytical solution of the attractor. The reason is that
the late time behavior of the field is not linear in potential of
$V(\phi)=V_{0}\exp(-\lambda{\phi}^2)$. Now, we will solve the
equation of motion of $\phi$ numerically.

\begin{figure}
\begin{center}
\includegraphics[width=2.5in]{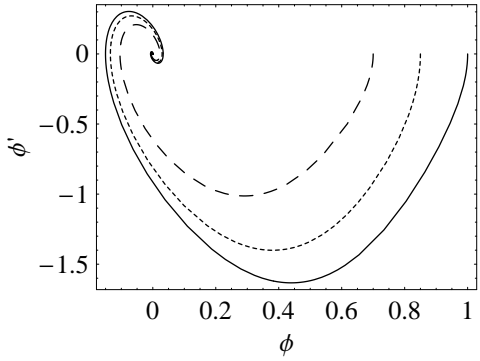}
\end{center}
\caption{The phase plane for different initial conditions, and
$\gamma_{m}=1$, $\lambda=2$}\label{plane}
\end{figure}

\begin{figure}
\begin{center}
\includegraphics[width=2.5in]{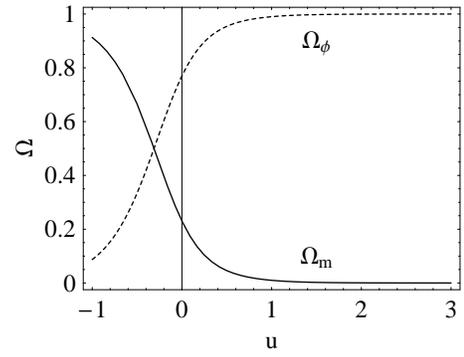}
\end{center}
\caption{The energy density parameter of DE and DM versus
$u=-\ln(1+z)$. The corresponding parameters are $\gamma_{m}=1$,
$\lambda=1$.}\label{omega}
\end{figure}

\begin{figure}
\begin{center}
\includegraphics[width=2.5in]{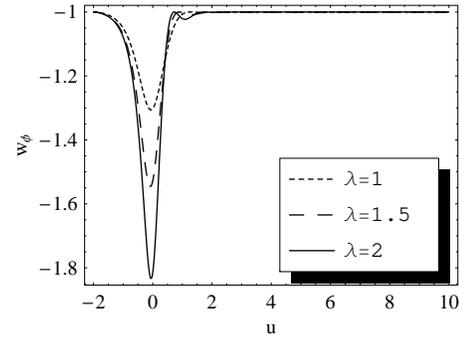}
\end{center}
\caption{The evolution of equation of state of phantom $w$ versus
$u=-\ln(1+z)$, in which $\gamma_{m}=1$, $\lambda=1$, $\lambda=1.5$
and $\lambda=2$.}\label{w}
\end{figure}

The numerical results show that there exists a stable attractor
solution which depends on $\lambda$ while is insensitive to the
initial conditions. In Fig.\ref{plane} we plot the
$(\phi,{\phi}^{'})$ phase diagram. It is shown in the phase plane
that the lines corresponding to different conditions will converge
together to the attractor solution with the cosmological evolution.
In Fig.\ref{omega}, we plot the evolution energy density parameters
of dark matter and dark energy with $\gamma_{m}=1$, $\lambda=1$. It
is exhibited that the dark energy occupies about $76\%$ and dark
matter occupies about $24\%$ at present, and the ratio between
energy densities of DE and DM will remain constant in the future. At
late time, the universe will be dominated by the phantom field
alone. From Fig.\ref{w} we can see that the parameter of equation of
state of phantom less than $-1$ at present, while it will approach
to $-1$ in the future.

\begin{figure}
\begin{center}
\includegraphics[width=2.5in]{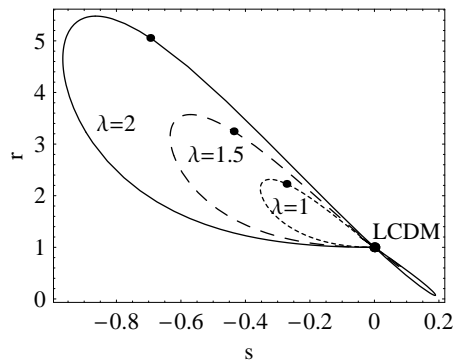}
\end{center}
\caption{The $s-r$ diagram of the phantom model. The curves evolve
in the variable interval $u\in[-2,5]$. Selected curves for
$\gamma_{m}=1$, $\lambda=1$, $\lambda=1.5$ and $\lambda=2$.
respectively. Dots locate the current values of the statefinder
parameters.}\label{sr}
\end{figure}

\begin{figure}
\begin{center}
\includegraphics[width=2.5in]{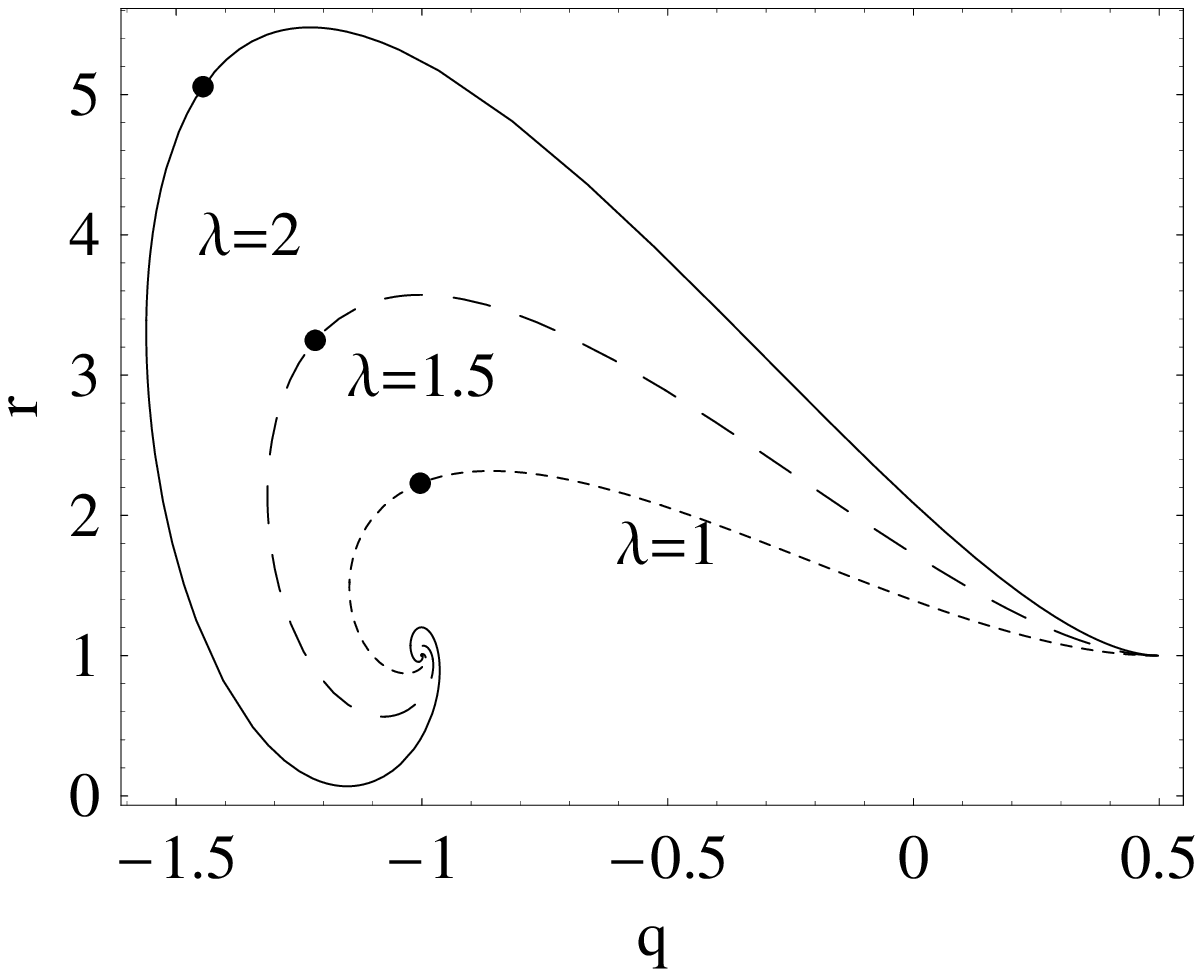}
\end{center}
\caption{The $q-r$ diagram of the phantom model. The curves evolve
in the variable interval $u\in[-2,5]$. Selected curves for
$\gamma_{m}=1$, $\lambda=1$, $\lambda=1.5$ and $\lambda=2$. Dots
locate the current values of the statefinder parameters.}\label{qr}
\end{figure}

The cosmological diagnostic pair $\{r,s\}$ called statefinder which
is introduced by Sahni et al. in \cite{statefinder} and defined as
\begin{equation}
r\equiv\frac{\dddot{a}}{aH^{3}},\qquad s\equiv\frac{r-1}{3(q-1/2)},
\end{equation}
Here $q$ is the deceleration parameter. The statefinder is a
"geometrical" diagnostic in the sense that it depends on the
expansion factor and hence on the metric describing space-time.
Since different cosmological models involving dark energy exhibit
qualitatively different evolution trajectories in the $s-r$ plane,
this statefinder diagnostic can differentiate various kinds of dark
energy models. For the spatially flat LCDM cosmological model, the
statefinder parameters correspond to a fixed point $\{r=1,s=0\}$. By
far some models, including the cosmological constant, quintessence,
phantom, quintom, the Chaplygin gas, braneworld models, holographic
models, interacting and coupling dark energy models
\cite{statefinder,my paper,models for statefinder}, have been
successfully differentiated. For example, the quintessence model
with inverse power law potential, the phantom model with power law
potential and the Chaplygin gas model all tend to approach the LCDM
fixed point, but for quintessence and phantom models the
trajectories lie in the regions $s>0, r<1$. We use another form of
statefinder parameters in terms of the total energy density $\rho$
and the total pressure $p$ in the universe:
\begin{equation}
r=1+\frac{9(\rho+p)}{2\rho}\frac{\dot{p}}{\dot{\rho}},\qquad
s=\frac{(\rho+p)}{p}\frac{\dot{p}}{\dot{\rho}}.
\end{equation}
Since the energy density of DE and DM is conserved, from (\ref{DE})
and (\ref{DM}) we can get:
\begin{eqnarray}
r&=&1-\frac{3}{2}w_{\phi}^{'}\Omega_\phi+\frac{9}{2}w_\phi(1+w_{\phi})\Omega_\phi,\\
s&=&1-\frac{w_{\phi}^{'}}{3w_\phi}+w_{\phi},
\end{eqnarray}
where $w_{\phi}^{'}=\frac{dw_{\phi}}{du}$. and the deceleration
parameter is also given
\begin{equation}
q=\frac{1}{2}(1+3w_{\phi}\Omega_\phi)
\end{equation}
In Fig.\ref{sr}, we show the time evolution of the statefinder pair
$\{r,s\}$. The plot is for variable interval $u\in[-2,5]$, and the
selected evolution trajectories of $r(s)$ correspond to
$\gamma_{m}=1$, $\lambda=1$, $\lambda=1.5$ and $\lambda=2$. We can
see that the $\{r,s\}$ exists in $s<0$ and $r>1$ and the
trajectories will pass through LCDM fixed point. It is interest to
find that the scope of $\{r,s\}$ would be largened with the
increasing of the $\lambda$, and in the future, the curves will
approach to the LCDM fixed point, which is quite different from
other phantom models with the exponential and pow law potential. We
also plot the statefinder pair $\{r,q\}$ in the Fig.\ref{qr}, in
which the corresponding parameters are the same as in the
Fig.\ref{sr}. We can see that the cosmic accerelation is ensured by
the phantom scalar field, and the curves will converge into a fixed
point in the future.

In summary, we investigate in this letter the attractor solution of
the phantom model with potential
$V(\phi)=V_{0}\exp(-\lambda{\phi}^2)$. The statefinder diagnostic
have be performed, which is shown that the evolving trajectories of
this scenario in the $s-r$ plane is quite different from other dark
energy models. We hope that the future high precision observation
will be capable of determining these statefinder parameters and
consequently shed light on the nature of dark energy.

\section*{ACKNOWLEDGMENTS}One of us (Chang Bao-rong) is grateful to Zhang Xin for helpful
discussion. This work was supported by NSF (10573003), NSF
(10647110), NBRP (2003CB716300) of P. R. China and DUT 893321.


\begin{thebibliography}{99}

\bibitem{Ia} A. G. Riess, {\it et al.} [Supernova Search Team
Collaboration] 1998 {\it Astron. J.} {\bf 116} 1009,
astro-ph/9805201; Perlmutter S 1999 {\it Astrophys. J.} {\bf 517}
565, astro-ph/9812133.

\bibitem{SDSS} A. C. Pope, {\it et. al} 2004 {\it
Astrophys. J.} \textbf{607} 655, astro-ph/0401249.


\bibitem{CMB} D. N. Spergel, {\it et. al}. 2003 \textit{Astrophys. J. Supp.}
\textbf{148} 175, astro-ph/0302209.



\bibitem{sahni9904398} V. Sahni, A. Starobinsky 2000 {\it Int. J. Mod. Phys. D} {\bf 9} 373-444, astro-ph/9904398.

\bibitem{cosmological constant} S. Weinberg 1989 {\it Rev. Mod. Phys.} {\bf 61} 1; S. M. Carroll 2001 {\it Living
Rev. Rel.} {\bf 4} 1, astro-ph/0004075; P. J. E. Peebles and B.
Ratra 2003 {\it Rev. Mod. Phys.} {\bf 75} 559, astro-ph/0207347;
 T. Padmanabhan 2003 {\it Phys. Rept.} {\bf 380} 235, hep-th/0212290.

\bibitem{quintessence1} R. R. Caldwell, R. Dave and P. J. Steinhardt 1998 {\it Phys.
Rev. Lett.} {\bf 80} 1582; M. S. Turner 2002 {\it Int. J. Mod. Phys.
A} {\bf 17S1} 180, astro-ph/0202008; V. Sahni 2002 {\it
Class.Quant.Grav.} {\bf 19} 3435, astro-ph/0202076.

\bibitem{quintessence2} I. Zlatev, L. Wang and  P. J. Steinhardt 1999 {\it Phys. Rev. Lett.}
{\bf 82} 896, astro-ph/9807002; P. J. Steinhardt, L. Wang, I. Zlatev
1999 {\it Phys. Rev. D} {\bf 59} 123504, astro-ph/9812313, X. Zhang
2005 {\it Mod. Phys. Lett. A} {\bf 20} 2575, astro-ph/0503072.


\bibitem{phantom} R. R. Caldwell, M. Kamionkowski,
 N. N. Weinberg 2003 {\it Phys. Rev. Lett.} {\bf 91} 071301,
astro-ph/0302506; P. Singh, M. Sami, N. Dadhich 2003 {\it Phys. Rev.
D} {\bf 68} 023522, hep-th/0305110.

\bibitem{astro-ph/0403228} Z. H. Zhu, M. K. Fujimoto and X. T. He 2004 {\it Astrophys. J.} {\bf 603}
365-370, astro-ph/0403228; L. Randall and R. Sundrum 1999 {\it Phys.
Rev. Lett.} {\bf 83} 3370; L. Randall and R. Sundrum 1999 {\it Phys.
Rev. Lett.} {\bf 83} 4690.


\bibitem{brane} Z. H. Zhu and J. S. Alcaniz 2005 {\it Astrophys. J.} {\bf 620}
7-11, astro-ph/0404201; G. Dvali, G. Gabadadze, M. Porrati 2000 {\it
Phys. Lett. B} {\bf 485} 208; C. Deffayet 2001 {\it Phys. Lett. B}
{\bf 502} 199, hep-th/0010186; C. Deffayet, G. Dvali and G.
Gabadadze 2002 {\it Phys. Rev. D} {\bf 65} 044023.


\bibitem{quattractor1} L. X. Xu,
 H. Y. Liu 2005 {\it Int. J.  Mod. Phys. D} {\bf 14} 1947-1957,
astro-ph/0507250; S. C. C. Ng, N. J. Nunes, F. Rosati 2001 {\it
Phys. Rev. D} {\bf 64} 083510; E. J. Copeland, A. R. Liddle and
 D. Wands 1998 {\it Phys. Rev. D} {\bf 57} 4686; B. R. Chang, H. Y. Liu and
 L. X. Xu 2005 {\it Mod. Phys. Lett. A} {\bf 20} 923, astro-ph/0405084.

\bibitem{guozk} Z. K. Guo, Y. S. Piao and Y. Z. Zhang 2004 {\it Phys. Lett.
B} {\bf 594} 247, astro-ph/0404225.

\bibitem{phattractor1} Z. K. Guo, Y. Z. Zhang 2005 {\it Phys.Rev. D} {\bf
71} 023501, astro-ph/0411524; Z. K. Guo, R. G. Cai and Y. Z. Zhang
2005 {\it JCAP} {\bf 0505} 002, astro-ph/0412624.

\bibitem{phattractor2} H. Y. Liu, H. Y. Liu, B. R. Chang and L. X. Xu 2005
{\it Mod. Phys. Lett. A} {\bf 20} 1973-1982, gr-qc/0504021.

\bibitem{statefinder} V. Sahni, T. D. Saini, A. A. Starobinsky and U. Alam 2003 {\it JETP Lett.} {\bf 77} 201
, astro-ph/0201498.

\bibitem{my paper} B. R. Chang, H. Y. Liu, L. X. Xu, C. W. Zhang and Y. L. Ping
 2007 {\it JCAP} {\bf 01} 016, astro-ph/0612616.

\bibitem{models for statefinder} V. Gorini, A. Kamenshchik and U. Moschella 2003 {\it
Phys. Rev. D} {\bf 67} 063509, astro-ph/0209395; X. Zhang 2005 {\it
Phys. Lett. B} {\bf 611} 1, astro-ph/0503075; X. Zhang 2005 {\it
Int. J. Mod. Phys. D} {\bf 14} 1597, astro-ph/0504586; P. X. Wu and
 H. W. Yu 2005 {\it Int. J. Mod. Phys. D} {\bf 14} 1873-1882,
gr-qc/0509036,
 X. Zhang, F. Q. Wu, J. F. Zhang 2006 {\it JCAP} {\bf 0601} 003.
\end{thebibliography}
\end{document}